\documentclass[journal]{IEEEtran}
\ifCLASSINFOpdf
\else
\fi
\usepackage[utf8]{inputenc} 
\usepackage[T1]{fontenc}    
\usepackage{hyperref}       
\usepackage{url}            
\usepackage{booktabs}       
\usepackage{amsfonts}       
\usepackage{nicefrac}       
\usepackage{microtype}      

\usepackage{graphicx}       
\usepackage{amsmath}        
\usepackage{amssymb}
\usepackage{verbatim}

\usepackage{helvet}  
\usepackage{courier}  
\usepackage{multirow}  

\usepackage{diagbox}  
\usepackage{color}  

\begin{document}
%
\title{Rapid Whole Slide Imaging via Learning-based Two-shot Virtual Autofocusing}
%
%
%

\author{Qiang~Li,
        Xianming~Liu,~\IEEEmembership{Member,~IEEE,}
        Kaige~Han,
        Cheng~Guo,
        Xiangyang~Ji,~\IEEEmembership{Member,~IEEE,}
        and~Xiaolin~Wu,~\IEEEmembership{Fellow,~IEEE}
\thanks{This work was supported by XX. (\emph{Corresponding author: Xianming Liu}).}

\IEEEcompsocitemizethanks{
\IEEEcompsocthanksitem Q. Li, X. Liu, K. Han are with the School of Computer Science and Technology, Harbin Institute of Technology, Harbin 150001, China, and also with the Peng Cheng Laboratory, Shenzhen 518052, China  (qiangli0120@hit.edu.cn; csxm@hit.edu.cn; 18S103171@stu.hit.edu.cn).
\IEEEcompsocthanksitem C. Guo is with the School of Measurement Science, Harbin Institute of Technology, Harbin 150001, China (guocheng\underline{ }27@163.com).
\IEEEcompsocthanksitem X. Ji is with the Department of Automation, Tsinghua University, Beijing 100084, China (xyji@tsinghua.edu.cn).
\IEEEcompsocthanksitem X. Wu is with the Department of Electrical and Computer Engineering,
McMaster University, Hamilton, ON L8S 4L8, Canada, and also with the
School of Electronic Information and Electrical Engineering, Shanghai Jiao
Tong University, Shanghai 200240, China (xwu@ece.mcmaster.ca).
}

\thanks{}}

%
%

\markboth{Journal of \LaTeX\ Class Files,~Vol.~14, No.~8, August~2015}%
{Shell \MakeLowercase{\textit{et al.}}: Bare Demo of IEEEtran.cls for IEEE Journals}
%



\maketitle

\begin{abstract}

Whole slide imaging (WSI) is an emerging technology for digital pathology. The process of autofocusing is the main influence of the performance of WSI. Traditional autofocusing methods either are time-consuming due to repetitive mechanical motions, or require additional hardware and thus are not compatible to current WSI systems. In this paper, we propose the concept of \textit{virtual autofocusing}, which does not rely on mechanical adjustment to conduct refocusing but instead recovers in-focus images in an offline learning-based manner. With the initial focal position, we only perform two-shot imaging, in contrast traditional methods commonly need to conduct as many as 21 times image shooting in each tile scanning. Considering that the two captured out-of-focus images  retain  pieces  of  partial  information  about  the  underlying in-focus image, we propose a U-Net-inspired deep neural network based approach for fusing them into a recovered in-focus image.  The proposed scheme is fast in tissue slides scanning, enabling a high-throughput generation of digital pathology images.
Experimental results demonstrate that our scheme achieves satisfactory refocusing performance.

\end{abstract}


%
\IEEEpeerreviewmaketitle

\section{Introduction}
%
%
%
%


\IEEEPARstart{W}{hole} 
slide imaging (WSI), also referred to as \emph{virtual microscopy} \cite{pantanowitz2011review,weinstein2009overview}, is applied to transform glass tissue slides to digital images. The interest in using WSI for digital pathology practice has steadily grown, since it provides a feasible approach
to assist disease diagnosis by convenient visualization and navigation of tissue slide images in an interactive manner \cite{higgins2015applications, abels2019computational}. 
A remarkable milestone is that in 2017 the US Food and Drug Administration has approved Philips' WSI system for the primary diagnostic use \cite{abels2017current}.

\begin{figure}[!t]
  \centering
  \includegraphics[width=1\linewidth]{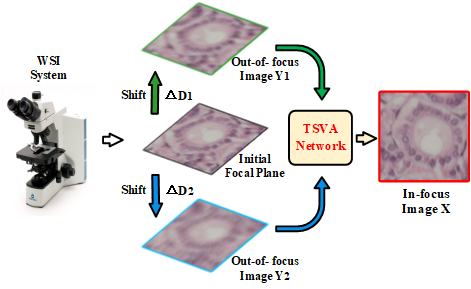}
  \caption{Illustration of the proposed rapid whole slide imaging system via learning-based two-shot virtual autofocusing.}
  \label{fig:model}
\vspace{-0.2cm}
\end{figure}

In a typical WSI system, different tiles of one tissue slide is scanned to obtain digital representations using a high-resolution objective lens, which are then aligned and stitched together to produce a complete and seamless image of the entire slide \cite{zarella2018practical}.  The numerical aperture (NA) of lens is usually high (typically larger than 0.75), and thus the resulting depth of field (DoF) is  micron-sized. The small DoF in WSI systems poses a challenge to acquire in-focus images of tissue sections that are with uneven topography \cite{langehanenberg2011autofocusing}. Thus the out-of-focus blurring artifact is the main cause of poor imaging quality in WSI \cite{kohlberger2019whole}. 
To address this problem, the most popular approach in current WSI systems is the focus map surveying method \cite{liao2017rapid}, which creates a focus map prior to scanning. Specifically, for each point in the map, it scans the sample to different axial positions and acquires a z-stack that include many out-of-focus images, which are further processed according to some criterion, such as image contrast or entropy, to derive the ideal focal point for one tile position. This process is repeated for all tiles of the entire tissue slide. The mechanical system then brings the sample back to the right position to perform in-focus image shooting.

There are two major limitations for this well-established focus map surveying method. First, as stated above, for each tile it overall needs $N+1$ times image shooting, where $N$ is the number of out-of-focus images in a z-stack and is usually set as $20$. The acquisition of a z-stack requires repetitive mechanical motions with cyclic acceleration and deceleration, which is time-consuming. Thus, creating a focus map for every tile requires a significant amount of overhead time. While selecting a subset of tiles for focus point surveying can save time to some extent, it comes at the expense of decreasing focusing accuracy and poor image quality. 
Second, since the derived ideal focal positions of different tiles are varying, focus map surveying puts forward higher requests to the mechanical system, which should have high positional accuracy and repeatability, in order to move the sample to the right position for the later scanning. Some hardware-assisted methods are presented to tackle the challenges listed above. For example, the setups of dual camera \cite{montalto2011autofocus} and two LEDs \cite{liao2016single} are proposed to reduce the number of axial scanning. However, the use of the additional hardware is not compatible with most existing WSI platforms. 

In this paper, we propose a novel whole slide imaging system with reduced scanning time cost and complete system compatibility.
Different from traditional autofocusing methods that rely on mechanical adjustment to conduct refocusing, we propose the concept of \textit{virtual autofocusing}, which instead recover in-focus images in a learning-based manner, as illustrated in Fig. \ref{fig:model}. Specifically, at the very beginning, for the first tile, we collect a z-stack with dense images, according to which we obtain the initial focal position. Then for the rest tiles, we perform two-shot imaging, where the z-stack contain only two out-of-focus images, which are captured in both sides of the initial focus with relative defocus offsets. This setup reduces the requirement of positional accuracy of the mechanical system, enabling a low-cost option. With these dual captured images, we no longer create the focus map and perform autofocusing during the process of tissue slide scanning, but recover the in-focus one offline with the help of large amounts of training data and high-performance computing devices such as GPU. In particular, considering the two-shot out-of-focus images retain pieces of partial information about the underlying in-focus image, we propose a U-Net-inspired deep neural network based approach for fusing them into a recovered in-focus image. Experimental results demonstrate that our scheme achieves satisfactory refocusing performance.

In summary, our scheme enjoys the following merits:
\begin{itemize}
  \item \textbf{High-speed and High-throughput}: Compared with traditional methods that conduct as many as 21 times image shooting in each tile scanning to create z-stack for focus map, our scheme only performs twice shooting instead of focus map surveying, which significantly reduces the overhead time cost. Moreover, our method does not perform autofocusing in the process of slide scanning, but recovers the underlying sharp image using an offline learning strategy. Therefore, it owns high scanning speed, which enables a high-throughput generation of digital pathology images.
    \item \textbf{High Imaging Quality}: The one-shot method has to exploit the additional constraints for the one out-of-focus image \cite{wu2019three}. More specifically, the single out-of-focus image contains complex features, resulting from the non-negligible sample thickness and inevitable optical aberrations. However, the proposed two-shot method utilizes more partial information about the underlying in-focus image from two out-of-focus images to fuse the features of mutual compensation. Therefore, the two-shot method has the higher imaging quality than the single one.
    \item \textbf{System Compatibility}: Our method does not involve any hardware modifications to current WSI systems. In contrast, other deep learning based autofocusing methods, such as Pinkard \textit{et al.} \cite{pinkard2019deep} and Jiang \textit{et al.} \cite{jiang2018transform}, need the addition of one or more LEDs to achieve refocusing based on two-shot images.

\end{itemize}

The paper is organized as follows: Section II overviews some related works. Section III introduces the problem formulation and the motivation of the proposed scheme, Section IV details the proposed in-focus image recovery method, and Section V provides the experimental results.  We conclude this paper in Section VI.

\section{Related Work}
\label{sec:related}

\begin{figure}[!t]
  \centering
  \includegraphics[width=0.9\linewidth]{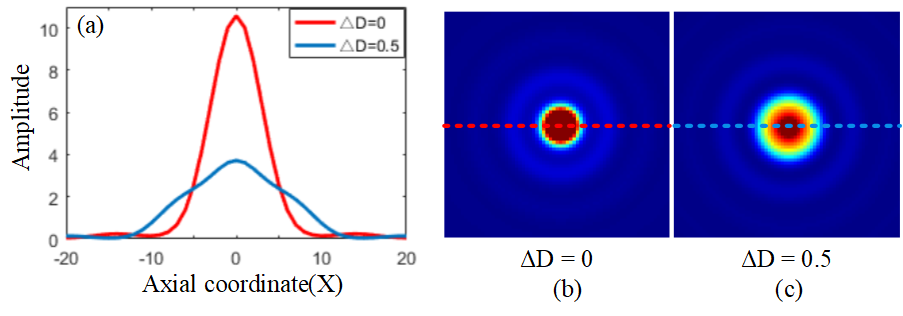}
  \caption{ (a) The axial PSF distribution curve with in-focus position (red line) and defocus position (blue line). (b) The lateral plane with $\Delta D=0$. (c) The lateral plane with $\Delta D=0.5$ $\mu$m.}
  \label{fig:PSF}

\vspace{-0.2cm}
\end{figure}

\begin{figure*}[!t]
  \centering
  \includegraphics[width=0.8\linewidth]{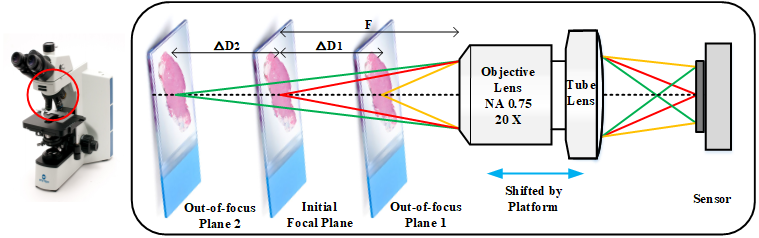}
 \vspace{0.2cm}
  \caption{Illustration of the microscopy imaging model of the proposed WSI system. }
  \label{fig:imaging}
\end{figure*}


In this section, we provide a brief review about existing autofocusing methods,
which can be generally divided into two categories: software-based and hardware-based. These two kinds of methods are sometimes combined to achieve satisfactory performance.

The basic idea of software-based autofocusing is to capture one or more out-of-focus images and use them to determine the ideal focal position. The most popular approach of this kind is focus map surveying \cite{liao2017rapid}. It captures a z-stack along the optical axis including a series of out-of-focus images with different relative distance offset \cite{subbarao1998selecting}, and then maximizes image contrast to determine the ideal focal plane. Software-based methods are usually slow due to the requirement of a full focal stack. Moreover, image contrast does not always serve as a good image quality metric. For example, some pathology samples that are weakly stained have low image contrast.

Hardware-based methods attempt to directly measure the distance from the objective lens to the sample, and thus is rapid.
Among hardware-based approaches, Liron \textit{et al.} \cite{liron2006laser} proposed to use an external light source or laser to measure the position of a reference point. 
Montalto \textit{et al.} \cite{montalto2011autofocus} proposed to utilize a secondary camera to decouple image acquisition from focusing and allow parallel processing.  Liao \textit{et al.}
\cite{liao2017rapid} developed a novel focus map surveying method using
additional LED illumination and autocorrelation image analysis. 
This kind of methods need to introduce hardware modifications to the microscope, which can be expensive and not compatible to current WSI systems.

With the emergence of deep learning in microscopy, convolutional neural networks (CNNs) based approaches have appeared for autofocusing.  The work in \cite{jiang2018transform} is the first one in the literature that uses CNN to predict the focal position. Specifically, Jiang \textit{et al.} \cite{jiang2018transform} firstly acquired $\sim$130,000 images with different defocus distances as the training dataset, and used an end-to-end deep residual network to build the relationship between the input image and its focal distance. This approach is able to capture images on the fly without focus map surveying. However, despite this method achieves remarkable autofocusing performance, from the perspective of methodology, it is not easy to derive a model that accurately describes the relationship between an image with complex contents and a numerical value (the defocus distance). More recently, Pinkard \textit{et al.} \cite{pinkard2019deep} proposed to combine the hardware modification and deep learning. It requires the addition of one or a few off-axis LEDs to a conventional transmitted light microscope. Defocus distance is then estimated and corrected based on a single image under this LED illumination using a neural network.


\section{Problem Formulation}
\label{sec:preliminaries}

Traditional autofocusing methods rely on mechanical adjustment to conduct refocusing, which need repetitive axial scanning. In order to reduce the time cost of scanning, we propose the concept of \textit{virtual autofocusing}, which no longer performs time-consuming mechanical autofocusing but instead recovers in-focus images in a learning-based manner. In this section, we introduce the problem formulation and the motivation of the proposed scheme.

\subsection{Optical Model}
In optical microscopy, the point spread function (PSF) can be formulated by the Born \& Wolf model \cite{born2013principles,hosseini2018focus}: 
\begin{equation}\label{PSF}
h(r, \Delta D)=\left|C \int_{0}^{1} J_{0}\left(k \frac{\mathrm{NA}}{n} r \rho\right) e^{-\frac{1}{2} i k \rho^{2} \Delta D\left(\frac{\mathrm{NA}}{n}\right)^{2}} \rho d \rho\right|^{2},
\end{equation}
where $r$ is the radial distance along the lateral plane; 
$\Delta D$ is the distance between the in-focus position and the imaging plane along the optical axis, \textit{i.e.}, the defocus distance;
$C$ is a normalization constant;
$J_{0}$ is zero-order Bessel function of the first kind;
$k$ is angular wave number of the light source;
$n$ is the refractive index;
$i$ is the imaginary number;
$\rho $ is the normalized coordinate in the exit pupil.
The axial PSF model is shown in Fig. \ref{fig:PSF} (a) and the lateral planes with different $\Delta D$ are shown in Fig. \ref{fig:PSF} (b) and (c). It can be found that, 
the amplitude of blue line ($\Delta D=0.5$ $\mu$m) is lower than the red one ($\Delta D=0$) due to the out-of-focus aberration, which becomes larger as $\Delta D$ increases.  

Accordingly, to recover the in-focus image, it is reasonable to assume that the most reliable knowledge is from the two nearest out-of-focus planes of the in-focus plane \cite{mcnally1999three, agard1984optical}.
In contrast, the single out-of-focus image has complex features which need to be restrained, due to the effects of sample thickness and optical aberrations.
Therefore, the two-shot method with feature fusion and compensation has the better performance than the single one.


\subsection{Infocusing and Defocusing Model}
In WSI, samples of pathological tissue slices are with uneven depth variations, and thus the ensuing PSFs vary spatially. 
Based on the layered depth of field model \cite{scofield1992212}, the continuous depth map is translated to discrete depth layers (image planes), and the PSF $h(r, \Delta D)$ is replaced by $h_{m}$, where $m$ stands for the position of each depth layer and $h_{0}$ is the PSF of the in-focus depth.
Each depth layer is blurred by its corresponding PSF with a convolution operation and the blurred depth layers are integrated to form the captured image. 
Therefore, the in-focus imaging model can be expressed as:
 \begin{equation}
 X = \sum_{m} x_{m} \otimes h_{m},
 \end{equation}
 where $x_{m}$ is the discrete depth layer of sample with depth $m$ and $x_{0}$ is the in-focus object plane of sample, $\otimes$ is the convolution operator, $X$ is the underlying in-focus image of $x_{0}$.
When the sample is shifted by offset $\Delta D_1$ from the in-focus object plane, we denote the new in-focus object plane as $x_{i}$. Similarly,  the new in-focus object plane is denoted as $x_{j}$ when the sample is shifted by offset $\Delta D_2$. 
Accordingly, the captured out-of-focus images $Y_{1}$ and $Y_{2}$ can be represented as:
\begin{equation}
Y_{1} = \sum_{m} x_{m+i} \otimes h_{m},\hspace{0.3cm}
Y_{2} = \sum_{m} x_{m+j} \otimes h_{m}.
\end{equation}

The defocusing imaging model indicates that recovering in-focus image from out-of-focus images in WSI is more challenging than the conventional inverse imaging problem. Fortunately, the two-shot out-of-focus images retain pieces of partial information about the underlying in-focus image, which inspires us to fuse them to derive the sharp image by a deep neural network.
\begin{figure*}
\begin{center}
\label{method_framwork}
\includegraphics[width=0.9\linewidth]{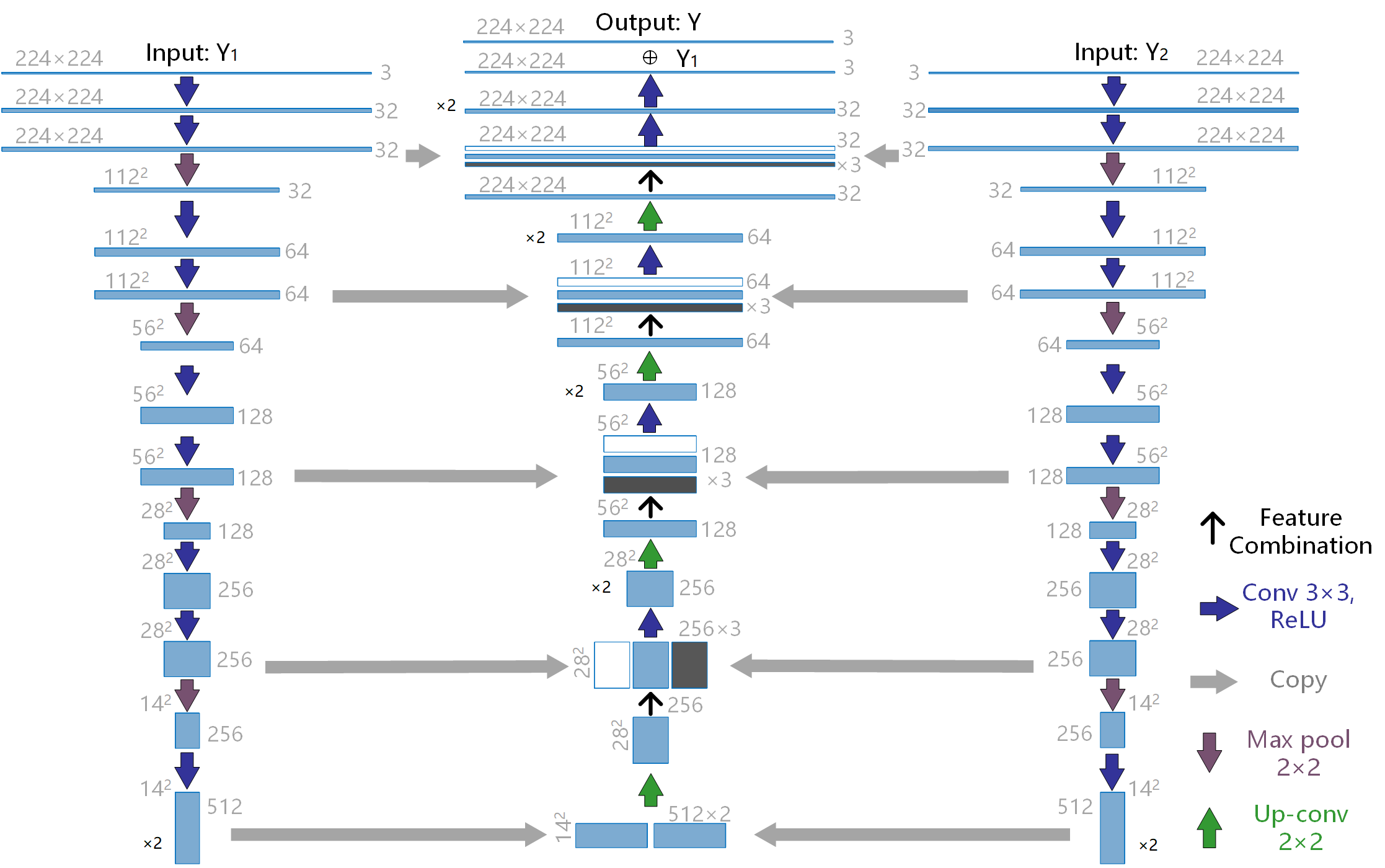}
\end{center}
\vspace{0.2cm}

   \caption{The architecture of the proposed TSVA network.
   Each blue box corresponds to a multi-channel feature map. 
The number of channels is denoted at the side edge of the box. 
The x-y-size is provided at the top edge of the box. 
White boxes represent copied feature maps of the left contracting path.
Black boxes represent copied feature maps of the right contracting path.
The colorful arrows denote the different operations. $\times$ 2 stands for an additional convolution.}

\label{fig:framework}
\vspace{-0.2cm}
\end{figure*}

\subsection{Virtual Autofocusing}
In view of the above, we propose a learning-based virtual autofocusing strategy relying on two-shot images, which are from the two nearest out-of-focus planes on both sides of the initial focal plane, as illustrated in Fig.~\ref{fig:imaging}. 
Specifically, at the very beginning, for the first tile we collect a full z-stack to derive the initial focal position $F$. It is worth noting that, since different tiles are with uneven topography, this position is usually not the focal one of other tiles. Then for the rest tiles, two out-of-focus images are captured with relative defocus offset $\Delta D_1$ and $\Delta D_2$ to $F$ respectively. This setting is inspired by the operation of manual microscopy, which first performs coarse tuning to get a best possible picture and further conducts fine tuning to finally acquire the best one. 

In conclusion, the practical workflow of the proposed method is listed as:
\begin{itemize}
  \item \textbf{Initial focal plane prediction:}
For the first tile, we collect a z-stack and obtain the initial focal position. 
\item \textbf{Two-shot imaging:}
For the rest tiles, we perform two-shot imaging, which are captured in both sides of the initial focal plane with relative defocus offsets.
\item \textbf{Algorithm processing:}
The in-focus image can be recovered directly offline by algorithm processing.

\end{itemize}

The following task is to recover $X$ by fusing its two observations $Y_{1}$ and $Y_{2}$. This is done in our scheme through a U-Net-inspired deep neural network, which will be elaborated in the next section. In practical implementation, we set $\Delta D_{1} = \Delta D_{2} = \Delta D$.


\section{The Proposed In-focus Image Recovery Method}
\label{sec:method}

With the dual captured images, we then try to recover the in-focus image with the help of large amounts of training data and high-performance computing environment. The proposed method is built upon an elegant deep neural network, the so-called U-net \cite{U-NET}. We modify and extend this architecture such that it can work with two input images and yield a recovered sharp image. In the following, we will introduce the network architecture and the training process in detail.

\begin{figure}[!t]
  \centering
  \includegraphics[width=0.9\linewidth]{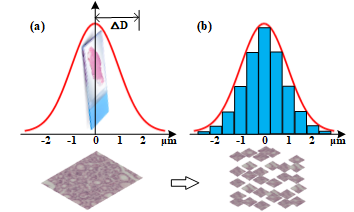}
  \caption{Illustration of Gaussian  distribution of focal positions. 
  (a) The Gaussian distribution of focal positions with $\Delta D$. The bottom tile shows the continuous fluctuations in the surface of the sample. (b) The discrete Gaussian distribution of focal positions with $\Delta D$. The bottom patches segmented from tiles exhibit the discrete offset of the sample.}
  \label{fig:distribution}

\vspace{-0.2cm}
\end{figure}


\begin{table*}
\scriptsize

\caption{Objective Performance Comparison with respect to PSNR (dB) of four compared methods.}
\centering
\vspace{0.20cm}
\begin{tabular}{p{2cm}p{0.8cm}<{\centering}ccccccc}
\toprule
Methods  & $\Delta D$  & Sample 1 & Sample 2 &  Sample 3& Sample 4&Sample 5&Sample 6 & Average \\

\hline

Dark Channel \cite{pan2016blind}   & \multirow{4}{*}{0.5$\mu$m} & 31.67 $\pm$1.69 & 29.92 $\pm$0.80 & 33.89 $\pm$0.88 & 33.56 $\pm$4.66 & 32.19 $\pm$3.10 & 30.30 $\pm$4.69 & 32.42 $\pm$3.85  \\ 

Graph Deblur \cite{bai2018graph}                                    & & 30.57 $\pm$15.59 & \textbf{51.76 $\pm$0.50} & 17.27 $\pm$3.87 & 22.85 $\pm$8.24 & 21.09 $\pm$3.85 & 22.97 $\pm$3.37  & 26.67 $\pm$12.79  \\ 

Burst U-net \cite{eccv18_burst}                                  & & 39.48 $\pm$1.26  & 38.83 $\pm$0.84  & 40.62 $\pm$0.26 & 39.44 $\pm$0.56 & 39.81 $\pm$0.94 & 40.41 $\pm$1.22 & 39.60 $\pm$0.95  \\ 

TSVA                                  &  & \textbf{44.83 $\pm$1.48} & 48.04 $\pm$0.40  &\textbf{48.35 $\pm$0.30}& \textbf{49.29 $\pm$1.00} & \textbf{49.61 $\pm$0.31}& \textbf{50.06 $\pm$0.54} & \textbf{48.71 $\pm$1.64}   \\\hline

Dark Channel \cite{pan2016blind}   & \multirow{4}{*}{1$\mu$m} & 30.74 $\pm$0.90 & 28.50 $\pm$0.05  & 32.76 $\pm$0.30 & 33.37 $\pm$1.16 & 30.95 $\pm$4.46 & 30.35 $\pm$2.77 & 31.74 $\pm$2.77    \\ 

Graph Deblur \cite{bai2018graph}                                    & & 28.91 $\pm$9.04 & 38.79 $\pm$0.33  & 17.98 $\pm$1.50 & 19.03 $\pm$3.42 & 17.38 $\pm$5.09 &  20.34 $\pm$4.09 & 21.78 $\pm$7.74   \\ 

Burst U-net \cite{eccv18_burst}                                    & & 36.37 $\pm$0.17 & 35.56 $\pm$0.03  &38.03 $\pm$0.04& 37.28 $\pm$0.26 & 37.70 $\pm$0.54& 38.50 $\pm$0.81 & 37.31 $\pm$0.89    \\ 

TSVA                                  &  & \textbf{37.13 $\pm$0.34} & \textbf{39.16 $\pm$0.34}  &\textbf{39.81 $\pm$0.24}& \textbf{40.37 $\pm$0.36} & \textbf{40.44 $\pm$0.16}& \textbf{41.35 $\pm$0.78}  & \textbf{39.93 $\pm$1.34}  \\\hline

Dark Channel \cite{pan2016blind}   & \multirow{4}{*}{1.5$\mu$m} & 30.35 $\pm$0.89 & 28.15 $\pm$1.28  & 33.65 $\pm$0& 32.99 $\pm$0.67 & 32.43 $\pm$0 & 31.88 $\pm$0.84 & 31.35 $\pm$1.73    \\ 

Graph Deblur \cite{bai2018graph}                                    & & 20.77 $\pm$8.46 & 37.50 $\pm$0.65  &10.62 $\pm$& 20.55 $\pm$1.33 & 19.17 $\pm$0 &  23.82 $\pm$4.49 & 22.19 $\pm$7.78  \\ 

Burst U-net \cite{eccv18_burst}                                     & & 34.27 $\pm$0.58 & 36.84 $\pm$0.20  &37.32 $\pm$0   & 36.80 $\pm$0.47 & 37.49 $\pm$0 & 36.92 $\pm$0.58  & 35.88 $\pm$1.40   \\ 

TSVA                                  &  & \textbf{35.18 $\pm$0.48} & \textbf{38.05 $\pm$0.56}  &\textbf{38.91 $\pm$0}& \textbf{39.09 $\pm$0.29} & \textbf{39.19 $\pm$0}& \textbf{39.77 $\pm$0.44} & \textbf{37.58 $\pm$2.02}   \\\hline

Dark Channel \cite{pan2016blind}   & \multirow{4}{*}{2$\mu$m} & 28.15 $\pm$0.52 & 28.06 $\pm$0.54 & 31.06 $\pm$0 & 32.31 $\pm$0.71 & 32.28 $\pm$0 & 30.92 $\pm$0  & 29.94 $\pm$1.95  \\ 

Graph Deblur \cite{bai2018graph}                                    & & 31.61 $\pm$0.67& 36.03 $\pm$0.26 &19.98 $\pm$0 & 18.77 $\pm$0.12 & 26.79 $\pm$0 & 24.10 $\pm$0 & 27.53 $\pm$6.47   \\ 

Burst U-net \cite{eccv18_burst}                                      & & 32.57 $\pm$0.20 & 35.23 $\pm$1.05 & 36.11 $\pm$0 & 36.10 $\pm$0.27 & 36.34 $\pm$0 & 36.86 $\pm$0   & 34.97 $\pm$1.71  \\ 

TSVA                                  &  & \textbf{33.01 $\pm$0.20} & \textbf{36.54 $\pm$0.21}  &\textbf{37.23 $\pm$0}& \textbf{37.85 $\pm$0.24} & \textbf{38.59 $\pm$0}& \textbf{39.84 $\pm$0} & \textbf{36.35 $\pm$2.38}   \\\hline

Dark Channel \cite{pan2016blind}   & \multirow{4}{*}{2.5$\mu$m} & 28.32 $\pm$0.64 & -  & - & - &31.53 $\pm$0& -  & 29.39 $\pm$1.60   \\ 

Graph Deblur \cite{bai2018graph}                                    & & 17.29 $\pm$1.06 & -  & - & - &20.25 $\pm$0 & -  & 18.27 $\pm$1.64   \\ 

Burst U-net \cite{eccv18_burst}                                    & & 30.80 $\pm$0.28 & - & - &- & 35.84 $\pm$0 & -  & 32.48 $\pm$2.39   \\ 

TSVA                                  &  & \textbf{31.30 $\pm$0.33} & -  & - & - & \textbf{38.13 $\pm$0}& -  & \textbf{33.58 $\pm$3.23}  \\\hline

Dark Channel \cite{pan2016blind}   & \multirow{4}{*}{3$\mu$m} &-& - &-&  -   & 30.94 $\pm$0& - & 30.94 $\pm$0   \\ 

Graph Deblur \cite{bai2018graph}                                    & & - & -  &-&  -  & 12.66 $\pm$0& - & 12.66 $\pm$0    \\ 

Burst U-net \cite{eccv18_burst}                                   & & - & -  &-&  -  & 34.86 $\pm$0.20 & - & 34.86 $\pm$0   \\ 

TSVA                                  &  & - & -  & - & - & \textbf{36.12 $\pm$0}  & - & \textbf{36.12 $\pm$0}  \\
\bottomrule
\label{table:test3}
\vspace{-0.9cm}
\end{tabular}
\end{table*}

\subsection{Network Architecture}

The proposed deep neural network that is tailored for two-shot virtual autofocusing (TSVA) is illustrated in Fig. \ref{fig:framework}.
Specifically, the TSVA network consists of two contracting paths (left and right sides) that are with two out-of-focus images $Y_1$ and $Y_2$ as inputs, and an expansive path (middle side) that outputs the recovered in-focus image $X$. The sharper image of two captured ones is chosen as $Y_1$, according to the metric of Brenner gradient \cite{sun2005autofocusing}.
\begin{itemize}
  \item \textbf{Contracting paths design:}
The contracting paths employ the typical convolutional architecture, including the
repeated use of two $3 \times 3$ convolutions followed by a rectied linear unit (ReLU) and $2 \times 2$ max pooling downsampling layer with stride 2. 
We double the number of feature channels at every downsampling step. 
These two paths share the same parameters.
Finally, we combine the deepest layers of two paths into a cascaded one.
\item \textbf{Expansive path design:}
The expansive path in each step includes an upsampling feature layer followed by $2 \times 2$ convolution (up-convolution), which halves the number of feature channels.
We build a concatenation with the corresponding feature maps from the left contracting path (white layer) and the right contracting path (black layer), and employ two $3 \times 3$ convolution followed by ReLU.
At the final residual layer, $Y_{1}$ is added to generate the recovered in-focus image $X$. 
In total the network has 27 convolutional layers.

\end{itemize}





\begin{figure*}[!t]
\begin{center}
\includegraphics[width=0.9\linewidth]{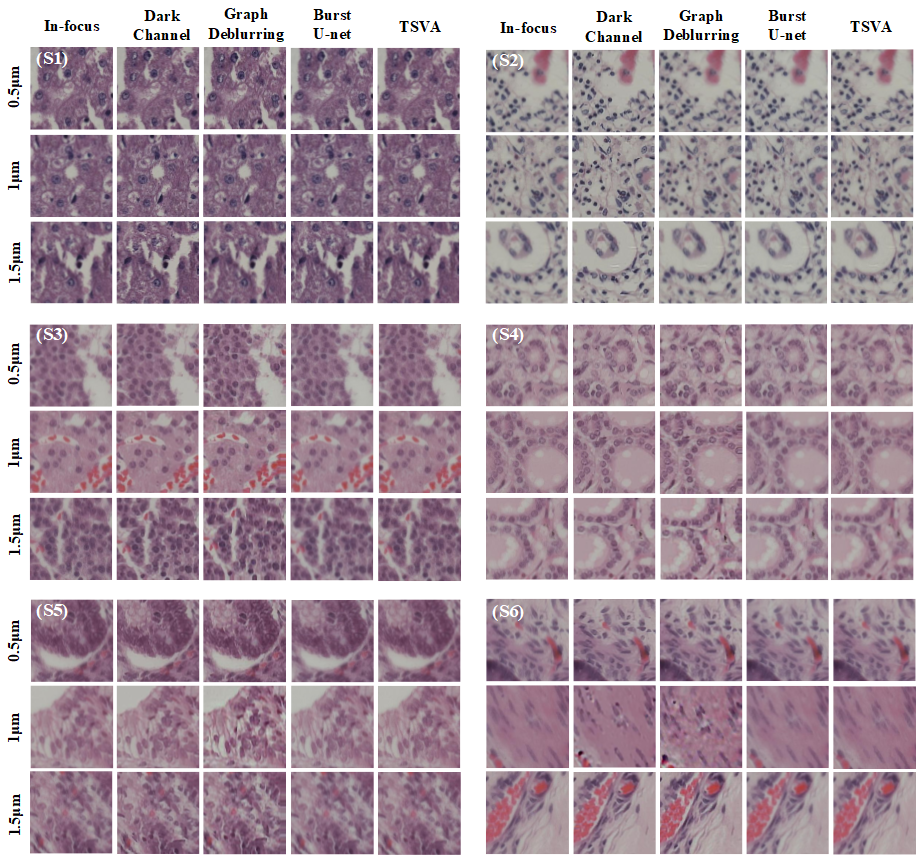}
\end{center}
   \caption{Subjective performance comparison on $Sample 1$ to $Sample 6$. Please enlarge the PDF for more details.}
\label{fig:test3}
\end{figure*}

\subsection{Network Training}

\subsubsection{Training Dataset}

We use a part of the dataset collected by Jiang \textit{et al.} \cite{jiang2018transform} to train our network. 
The dataset includes 35 research-grade human pathology slides with Hematoxylin and eosin stains (Omano OMSK-HP50), and contains 162 pathological tissue z-stack tiles.  
For each tile there is a stack of 41 images taken with different focal distances in a step size of 0.5$\mu m$, ranging from -10$\mu m$ to 10$\mu m$, with 0$\mu m$ corresponding to the image in focus.  
The in-focus image is recovered by maximizing Brenner gradient of the z-stack images. 

In image stacks of all tiles, the focal distance of an out-of-focus image is given as the defocus offset to the image in focus.
But in our system, the microscope camera makes two shots of each title at two prefixed focal distances.
Therefore, we need the out-of-focus images of absolute focal distances to train our TSVA network.  
We convert the training images in relative focal distance in the dataset of \cite{jiang2018transform} to those in absolute focal distance by simply adding a Gaussian random variable $n\sim{\cal N}(0,1)$ to the relative focal distance.  
This is because, according to the observation of \cite{hart2014focal}, the focal positions follow a Gaussian distribution, as shown in Fig. \ref{fig:distribution} (a).
Specifically, The images of slides are divided into $224 \times 224$ patches in Fig. \ref{fig:distribution} (b).
Then, we convert the dataset to discrete patches of Gaussian distribution. 
There are 3240 patches in the initial dataset and we enlarge the dataset by rotation.

\subsubsection{Implementation Details}

Here we clarify some details in implementation. 
In network training, the loss function is defined as follows: 
\begin{equation}
L=\frac{1}{N}\sum_{i=1}^{N}(X_{i}-\tilde{X_{i}})^{2},
\end{equation}
where $X_{i}$ is the ground-truth in-focus image and $\tilde{X_{i}}$ is the network output, and $N$ is the number of training images in each batch.
We select 85\% patches with labeled relative defocus offset $\Delta D$ as our training set and 15\% patches for verification.
We utilize batch normalization with batch size as 20 for acceleration training.
The network is trained using the ADAM optimizer with a learning rate as 0.0005 for 50 epochs. 
The network training is run on a single NVIDIA GTX 1080Ti.

\begin{figure*}[!t]
  \centering
  \includegraphics[width=0.9\linewidth]{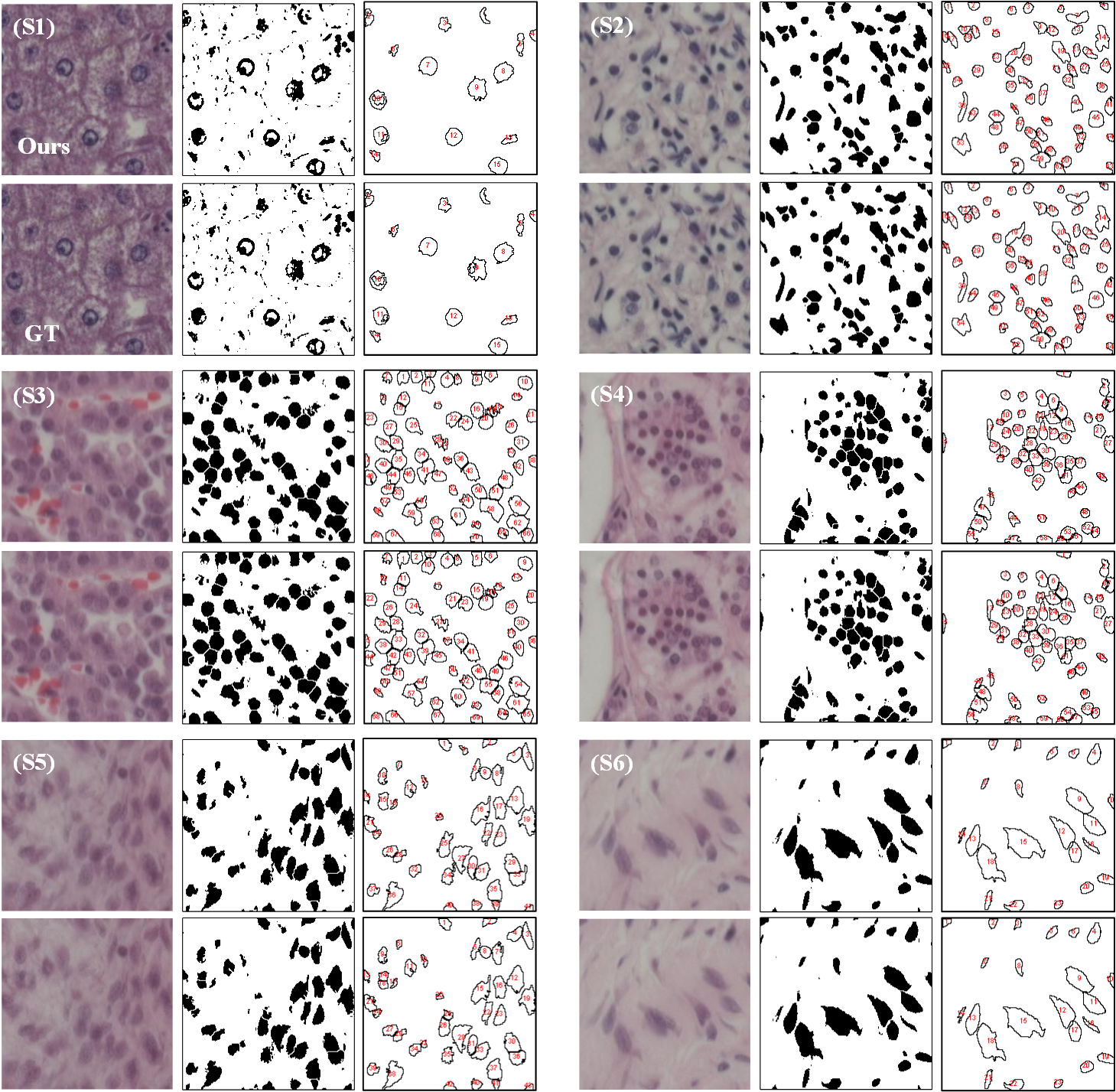}\\
  \vspace{0.20cm}
  \caption{Influence of image quality to the accuracy of cell counting. For (S1) to (S6), the cell counting results on our generated image are at the top and the corresponding results of ground-truth are at the bottom. From left to right, the input image for cell counting, the cell segmentation image, and the image of cell outlines counting.  Please enlarge the PDF for more details.}
  \label{fig:cell-counting}
\end{figure*}

\begin{figure*}[!t]
  \centering
  \includegraphics[width=0.9\linewidth]{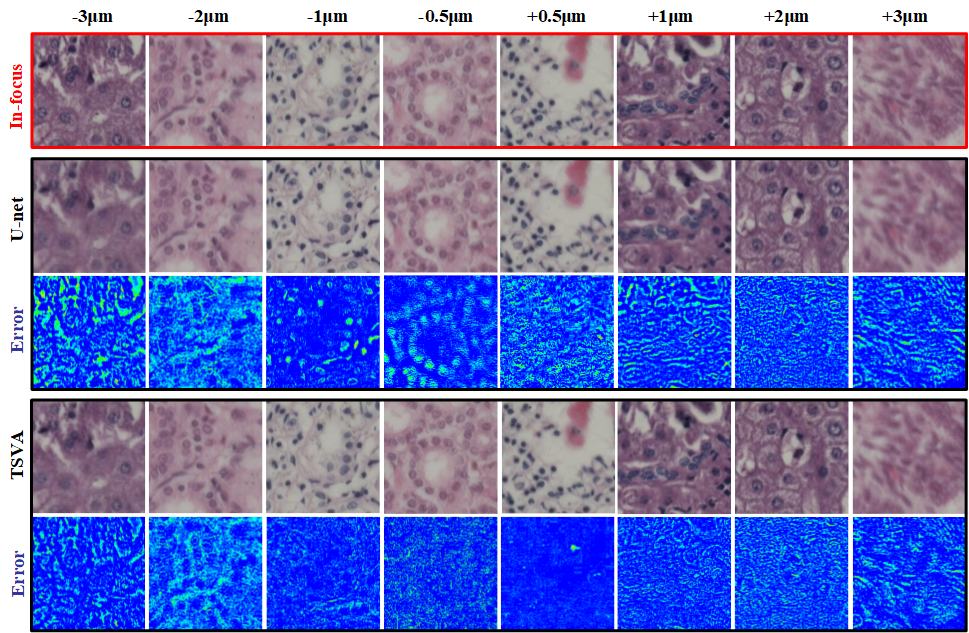}\\
  \vspace{0.20cm}
  \caption{Subjective Performance Comparison on images of Dataset 1. Please enlarge the PDF for more details. The in-focus images in red block are ground truth. The results of U-net and TSVA are shown in the two black blocks with the corresponding error maps on the bottom. }
  \label{fig:test1_1}
\end{figure*}

\begin{table*}
\scriptsize

\caption{The average numbers of counted cells with respect to different $\Delta D$ on all samples in Dataset 1.}
\centering
\vspace{0.20cm}
\begin{tabular}{p{0.8cm}cc|cc|cc|cc|cc|cc|cc}
\toprule
\multirow{2}{*}{$\Delta D$}    & \multicolumn{2}{c|}{Sample 1} & \multicolumn{2}{c|}{Sample 2}&  \multicolumn{2}{c|}{Sample 3}& \multicolumn{2}{c|}{Sample 4}&\multicolumn{2}{c|}{Sample 5}&\multicolumn{2}{c|}{Sample 6} &  \multicolumn{2}{c}{Average} \\
 \cline{2-15}                             &Ours&GT&Ours&GT&Ours&GT&Ours&GT&Ours&GT&Ours&GT&Ours&GT\\
\hline

0.5$\mu$m & 15.17& 15.5   & 42.38  & 42.13 & 22.4 & 22.4  & 19.4  & 19.36 & 18.75 & 19    & 22.75 & 23    &22.38 &22.43  \\ 
1$\mu$m   & 12.83& 13.16  & 39.67  & 40    & 18.5 & 19.5  & 21.12 & 21.18 & 18.43 & 18.86 & 18.17 & 17.83 &20.25 &20.42  \\ 
1.5$\mu$m & 14.44& 14.56  & 32     & 33    & 33   & 33    & 25.5  & 25.67 & 21    & 21    & 25    & 25.5  &21.78 &22.05  \\ 
2$\mu$m   & 15.33& 16     & 34.5   & 34    & 29   & 28    & 21    & 22    & 23    & 22    & 18    & 17    &22.70 &22.7  \\ 
2.5$\mu$m & 10   & 10     & -      & -     & -    & -     & -     & -     & 28    & 27    & -     & -     &16    &15.67  \\ 
3$\mu$m   & -    & -      & -      & -     & -    & -     & -     & -     & 27    & 29    & -     & -     &27    &29  \\ \hline
Average   & 14.00& 14.27  & 39.40  & 39.40 & 23.44& 23.56 & 20.78 & 20.84 & 19.70 & 19.96 & 21.20 & 21.20 &21.57 &21.69  \\ 
\bottomrule
\label{table:cell-counting}
\vspace{-0.9cm}
\end{tabular}
\end{table*}

\begin{figure*}[!t]
  \centering
  \includegraphics[width=0.9\linewidth]{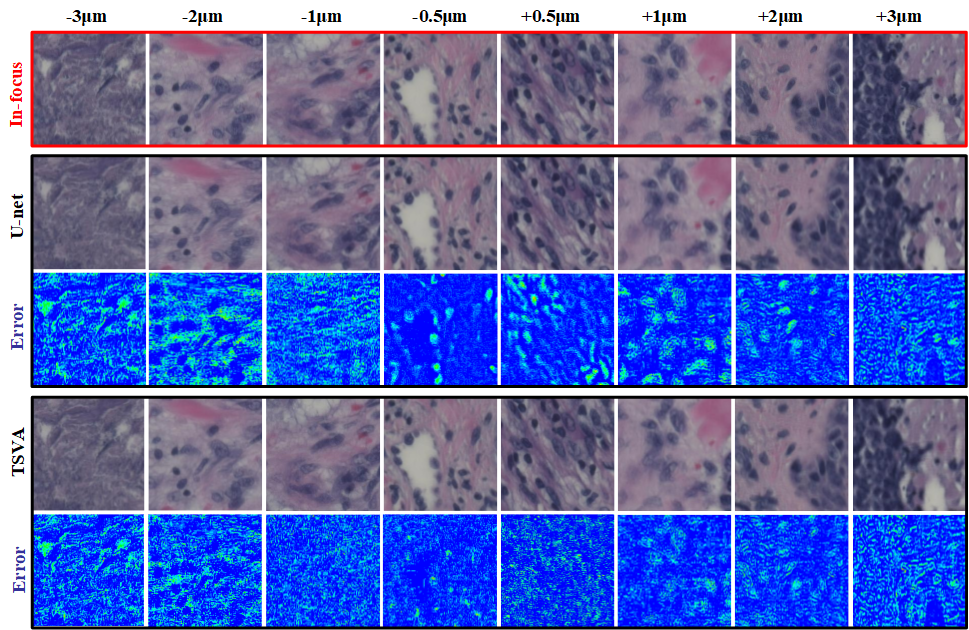}\\
  \vspace{0.20cm}
  \caption{Subjective Performance Comparison on images of Dataset 2. Please enlarge the PDF for more details. The in-focus images in red block are ground truth. The results of U-net and TSVA are shown in the two black blocks with the corresponding error maps on the bottom. }
  \label{fig:test1_2}
\end{figure*}


\begin{table*}[!h]
\scriptsize

\caption{PSNR performance comparison of U-net and TSVA on Dataset 1 and Dataset 2 with respect to different $\Delta D$.}
\centering
\vspace{0.20cm}

\begin{tabular}{p{0.8cm}<{\centering}p{0.8cm}<{\centering}cccccccc}
\toprule
Dataset&Methods& \multicolumn{7}{c}{Relative Distance Offset $\Delta D$ (The mean on the top and standard deviation (SD) on the bottom in each methods)  }&Average\\
\hline
\multirow{8}{*}{Dataset 1}&\multirow{4}{*}{U-net} & $\Delta D$ &-3$\mu$m&-2.5$\mu$m&-2$\mu$m&-1.5$\mu$m&-1$\mu$m&-0.5$\mu$m & \multirow{4}{*}{39.44 $\pm$3.76}\\

&&PSNR & 27.96 $\pm$0& 33.46 $\pm$4.17 & 38.28 $\pm$1.28 & 37.42 $\pm$2.86 & 38.86 $\pm$1.97 & 41.61 $\pm$4.08\\ 
                       \cline{3-9} 

&&0$\mu$m&+0.5$\mu$m&+1$\mu$m&+1.5$\mu$m&+2$\mu$m&+2.5$\mu$m&+3$\mu$m\\

&&39.44 $\pm$1.58 & 41.88 $\pm$4.36&  38.74 $\pm$2.26& 36.05 $\pm$3.11& 36.11 $\pm$2.78& 32.28 $\pm$3.90& 34.88 $\pm$0 & \\ 
                       \cline{2-10} 
          

&\multirow{4}{*}{TSVA}    & $\Delta D$ &-3$\mu$m&-2.5$\mu$m&-2$\mu$m&-1.5$\mu$m&-1$\mu$m&-0.5$\mu$m & \multirow{4}{*}{\textbf{42.25 $\pm$4.90}}\\

&& PSNR &  \textbf{30.11 $\pm$0} & \textbf{33.58 $\pm$2.74} & \textbf{38.60 $\pm$1.07} & \textbf{38.34 $\pm$1.87} & \textbf{39.82 $\pm$1.11} & \textbf{47.99 $\pm$1.59} \\
                      \cline{3-9}
                      
&&0$\mu$m&+0.5$\mu$m&+1$\mu$m&+1.5$\mu$m&+2$\mu$m&+2.5$\mu$m&+3$\mu$m\\
                        
&&\textbf{39.61 $\pm$1.32}& \textbf{48.71 $\pm$1.64} & \textbf{39.93 $\pm$1.34} & \textbf{37.58 $\pm$2.02} & \textbf{36.35 $\pm$2.38} & \textbf{33.58 $\pm$3.23} & \textbf{36.12 $\pm$0 } &  \\ 
\hline\hline
               

\multirow{8}{*}{Dataset 2}& \multirow{4}{*}{U-net}   & $\Delta D$ &-3$\mu$m&-2.5$\mu$m&-2$\mu$m&-1.5$\mu$m&-1$\mu$m&-0.5$\mu$m & \multirow{4}{*}{38.83 $\pm$2.95}\\

&&PSNR&  33.22 $\pm$0.34 & 34.82 $\pm$1.74 & 36.27  $\pm$1.65  & 37.76  $\pm$1.29& 38.61  $\pm$0.99 & 41.28  $\pm$3.23& \\
           \cline{3-9}

&&0$\mu$m&+0.5$\mu$m&+1$\mu$m&+1.5$\mu$m&+2$\mu$m&+2.5$\mu$m&+3$\mu$m\\

&&38.80  $\pm$0.97 & 40.44  $\pm$3.68 &  37.71  $\pm$1.54  & 36.15  $\pm$1.66 & 35.31  $\pm$1.90 & 33.49  $\pm$1.81 & 33.00  $\pm$1.12& \\
\cline{2-10}

                        
&\multirow{4}{*}{TSVA}   &$\Delta D$ &-3$\mu$m&-2.5$\mu$m&-2$\mu$m&-1.5$\mu$m&-1$\mu$m&-0.5$\mu$m & \multirow{4}{*}{\textbf{42.32 $\pm$4.67}}\\

&&PSNR& \textbf{34.87 $\pm$0.45} & \textbf{35.61 $\pm$1.20} & \textbf{37.20  $\pm$1.04} & \textbf{38.65  $\pm$0.63} &\textbf{ 39.78  $\pm$0.62}& \textbf{48.39 $\pm$0.91} \\
           \cline{3-9}

&&0$\mu$m&+0.5$\mu$m&+1$\mu$m&+1.5$\mu$m&+2$\mu$m&+2.5$\mu$m&+3$\mu$m\\

&&\textbf{39.55 $\pm$0.57} & \textbf{48.55 $\pm$0.88} &\textbf{39.77 $\pm$0.60}& \textbf{38.26 $\pm$0.79} & \textbf{36.61 $\pm$1.32} & \textbf{35.23 $\pm$0.79} & \textbf{33.89 $\pm$0.00} &  \\ 
                
\bottomrule
\label{table:test1_2}
\end{tabular}

\end{table*}



\section{Experiments}
\label{sec:experiments}

In this section, 
we provide extensive experimental results to demonstrate the effectiveness of our proposed TSVA scheme. 

The experimental analysis is conducted on two test datasets: 
\begin{itemize}
\item \textbf{Dataset 1:} We use the part of Dataset 1 \cite{jiang2018transform} except that for training as the test set. It contains all stained tissue slide images, including six categories of biological tissues with different morphological characteristics of size, thickness and structure, named $Sample 1$ to $Sample 6$. 

\item \textbf{Dataset 2:} Dataset 2 \cite{jiang2018transform} that contains the de-identified HE skin-tissue slides made by the Dermatology Department of the UConn Health Center is also used for testing, which is collected from different source with the training set. It includes seven categories of biological tissues named $Sample 7$ to $Sample 13$.
\end{itemize}

For both datasets, the size of each tile image is $1224\times1024$. 
We select test images with the corresponding relative defocus offset $\Delta D$ ranging from -3$\mu m$ to +3$\mu m$ with interval 0.5$\mu m$, which are also converted in the same way as the training data.
There are 340 and 640 patches in Dataset 1 and Dataset 2 respectively.

\begin{figure}[!t]
  \centering
  \includegraphics[width=0.9\linewidth]{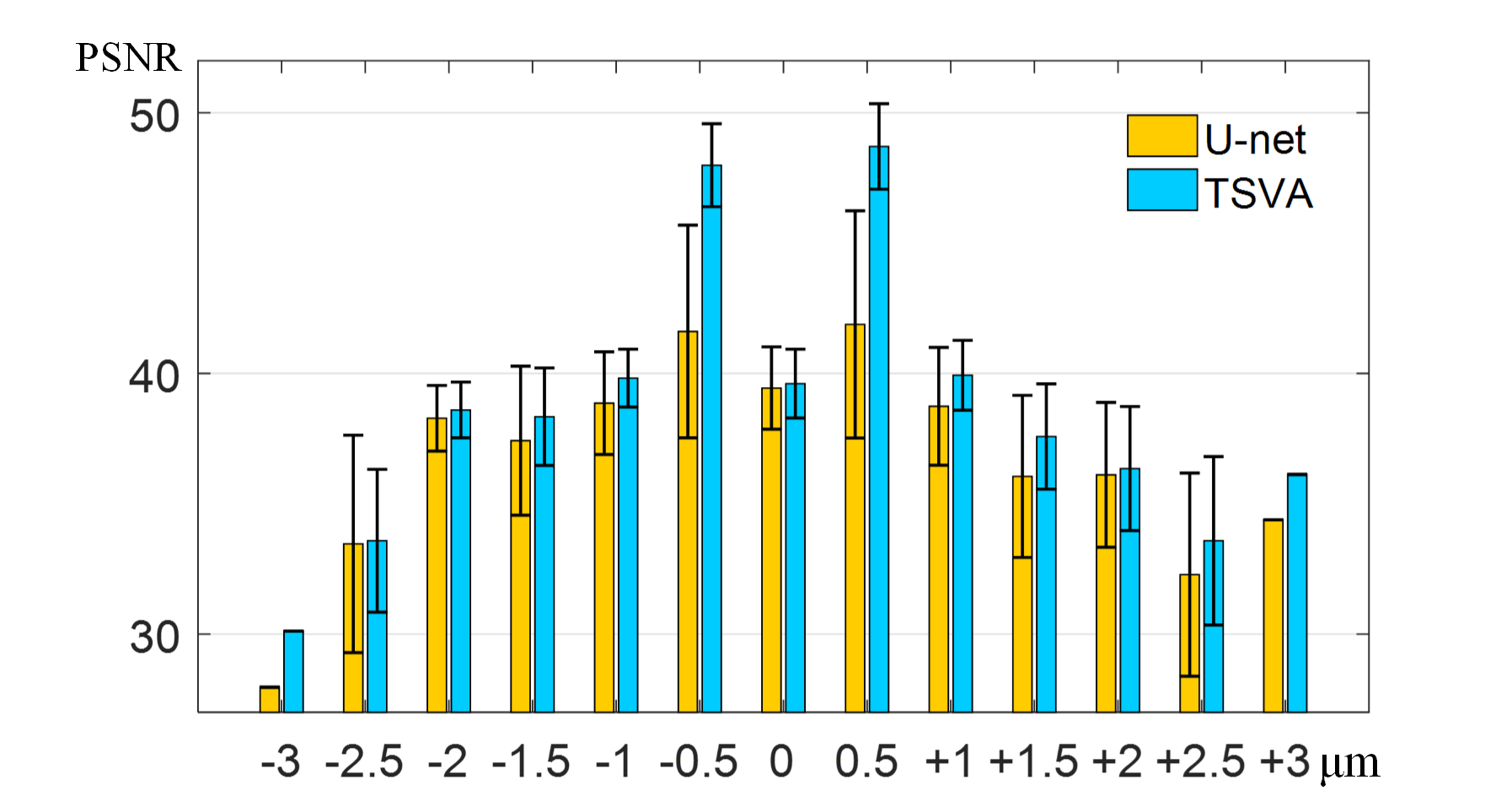}\\
 \vspace{0.2cm}
  \caption{PSNR performance comparison of U-net and TSVA on Dataset 1 with respect to different $\Delta D$. }
  \label{fig:h1}
 \vspace{-0.4cm}

\end{figure}

\begin{figure}[!t]
  \centering
  \includegraphics[width=0.9\linewidth]{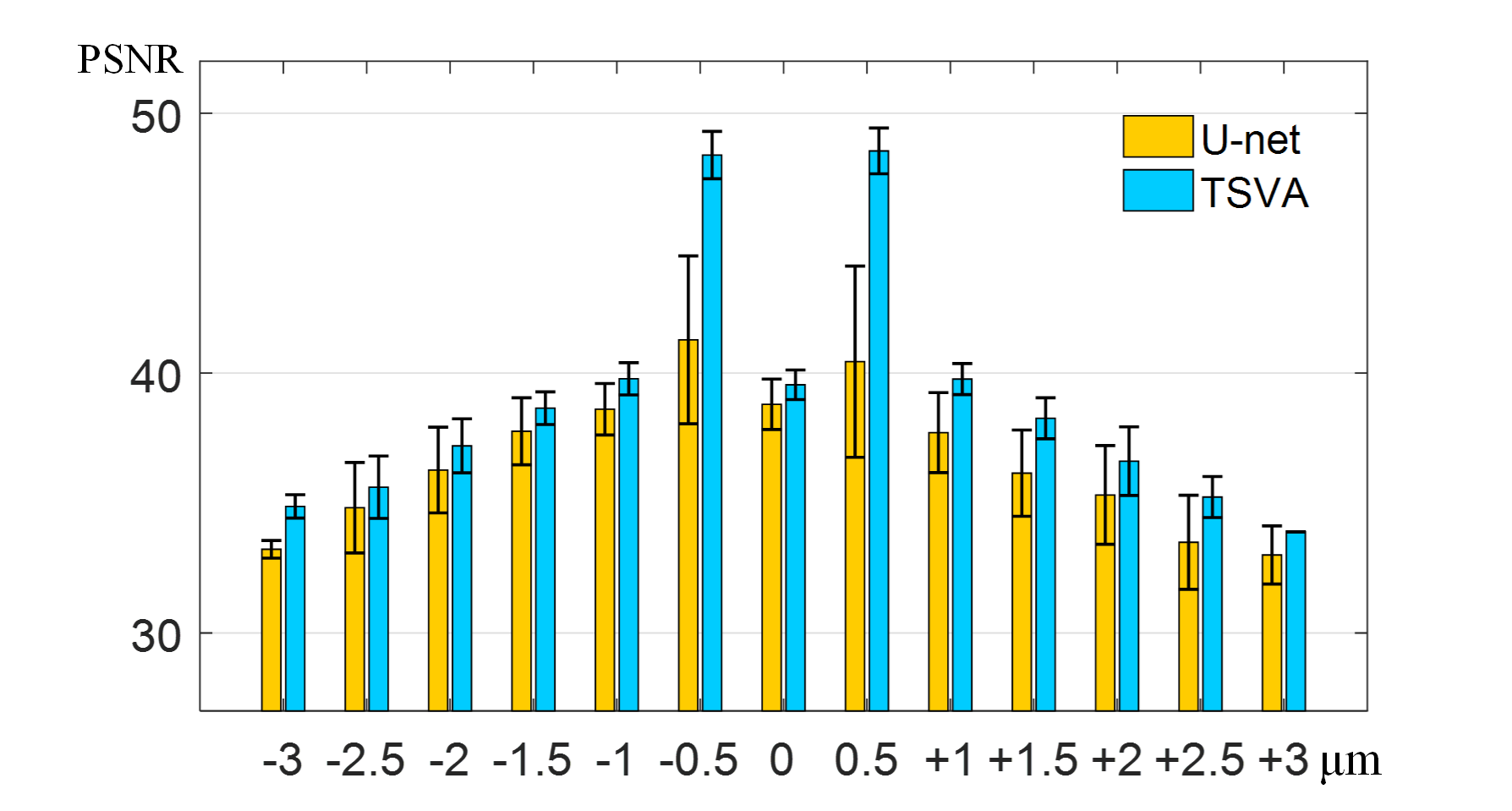}\\
 \vspace{0.2cm}
  \caption{PSNR performance comparison of U-net and TSVA on Dataset 2 with respect to different $\Delta D$. }
  \label{fig:h2}
 \vspace{-0.4cm}

\end{figure}

\subsection {Comparison with State-of-the-arts }

In this subsection, to demonstrate the effectiveness of the proposed in-focus image recovery scheme, we provide objective and subjective quality comparison on Dataset 1 with state-of-the-art image deblurring methods, including dark channel prior based \cite{pan2016blind}, graph-prior based \cite{bai2018graph}, U-net based burst deblurring \cite{eccv18_burst} that also takes multiple images as inputs. 

The objective performance evaluation with respect to PSNR is shown in Table \ref{table:test3}, where ``-" represents there is no corresponding image in this defocus distance.
It can be found that, our method achieves the best PSNR performance on all sample images.
These comparison results demonstrate the superior performance of our proposed TSVA network.

The subjective comparison results on six test images are illustrated in Fig. \ref{fig:test3}.
The ground-truth in-focus images are also offered as the quality reference. From the results, it can be found that the statistical prior-based methods, \textit{i.e.}, \cite{pan2016blind} and \cite{bai2018graph}, cannot handle complicated out-of-focus effects in WSI, since the statistics of biomedical images is different from natural images. These two methods cannot preserve texture information well. Burst U-net based method \cite{eccv18_burst}, which also uses deep neural network for burst deblurring, achieves better subjective performance than \cite{pan2016blind} and \cite{bai2018graph}. Our method achieves the best subjective performance among compared methods. The recovered in-focus images share a very close subjective effect with the ground-truth in-focus images.



\subsection{Influence of Image Quality to Downstream Image Analysis}

According to Fig. \ref{fig:test3}, it is hard to differentiate the recovered in-focus images from the ground-truth by human eyes.
Another concern is whether the machine also cannot differentiate them, \textit{i.e.}, whether the recovered in-focus images would significantly reduce the accuracy of downstream image analysis tasks?

In this subsection, using cell counting that is a typical task of pathology image analysis as an example, we examine the influence of the quality of images yielded by the proposed visual autofocusing approach to the counting accuracy.
We utilize a widely used tool \textit{ImageJ}\footnote{https://imagej.nih.gov/ij/} \cite{schneider2012nih} released by National Institutes of Health (NIH) as the test platform, which conducts cell counting including the following four steps: 1) gray processing; 2) adjusting brightness and contrast; 3) thresholding; 4) analysis of cell counting.

The in-focus images recovered by our method and the ground-truth in-focus images are taken as input to \textit{ImageJ}, respectively.  The results of cell counting are illustrated in Fig. \ref{fig:cell-counting}.
It can be found that, the cell counting results on our recovered images are very close to the results on the corresponding ground-truth images. 
In Table \ref{table:cell-counting}, we also show the comparison of numbers of counted cells with respect to different $\Delta D$ on $Sample 1$ to $Sample 6$. It can be seen that, compared with the results on the ground-truth (GT), the average cell counting error on our recovered in-focus images is 0.12, which is too small to reduce the accuracy of downstream analysis significantly.

\subsection{Ablation Study}

In this subsection, we provide the empirical ablation analysis of the proposed TSVA network.
According to the TSVA architecture, there are two input out-of-focus images with relative defocus offsets $\Delta D$.
Therefore, it is essential to analyze the influence of dual input images and relative defocus offsets to the final performance.  Moreover, we provide the study of the robustness of the proposed scheme to different test sets. Considering that our TSVA is built upon the U-net, we employ the traditional U-net \cite{U-NET} as the baseline, which takes $Y_{1}$ as input with different relative distance offsets.

\subsubsection{Influence of dual input images} In this part, we provide empirical analysis if the dual captured images is really helpful to improve the quality of recovered in-focus images compared with the single one.

Table \ref{table:test1_2} shows objective performance comparison of U-net that takes $Y_{1}$ as input and our TSVA that takes $Y_{1}$ and $Y_{2}$ as inputs. 
It can be found that, on Dataset 1 and 2, TSVA achieves better PSNR performance than U-net for all cases. 
The average PSNR gains are 2.81dB and 3.49dB over U-net, respectively. 

We also provide subjective performance comparison of U-net and TSVA in Fig. \ref{fig:test1_1} on Dataset 1.
For clear display,  we show the error maps between the recovered in-focus images and the corresponding ground-truth. It can be seen that, compared with U-net, the structure errors produced by TSVA are smaller, in particular when $\Delta D$ is ranging from  from -1$\mu m$ to +1$\mu m$.
Therefore, the proposed TSVA network achieves superior performance than U-net, benefiting from the dual inputs. 


\subsubsection{Influence of different relative defocus offsets}
In this part, we examine the influence of different relative defocus offsets to the final performance.

The PSNR histograms with respect to $\Delta D$ on Dataset 1 and Dataset 2 are shown in Fig. \ref{fig:h1} and Fig. \ref{fig:h2} respectively.
It can be found that:
i) For different $\Delta D$, the proposed TSVA always achieves higher PSNR values than U-net. 
This demonstrate that the performance of our scheme is robust with respect to $\Delta D$.
ii) The highest PSNR gains appear when $\Delta D=+0.5$ $\mu$m and $\Delta D=-0.5$ $\mu$m.
In practical case, most of estimated focal positions also lie in the region of $\pm$ 0.5 $\mu$m.
Therefore, the TSVA network realizes virtual autofocusing with high accuracy.

\subsubsection{Influence of different test sets} In this part, we examine the robustness of our method to different test sets. In Table \ref{table:test1_2}, we provide objective performance evaluation with respect to PSNR on samples of Dataset 2.
It can be found that, for test samples from different resources of the training set, our method still achieves the best PSNR performance for all cases. The average PSNR gain over U-net is 3.49dB. 
The subjective performance comparison of U-net and TSVA is shown in Fig. \ref{fig:test1_2} on Dataset 2. Similar to the results on Dataset 1, the structure errors produced by TSVA is also much smaller than U-net.
These results demonstrate that the proposed TSVA network has a strong generalization capability.

\begin{table}
\scriptsize

\caption{The workflow comparison between the conventional methods and the proposed method.}
\centering
\vspace{0.20cm}
\begin{tabular}{ccc}
\toprule
Step    & Conventional Methods & Proposed Method \\
\hline

(a)  & \emph{Create a z-stack for the first tile} & \emph{Create a z-stack for the first tile}    \\ 
(b)  & \emph{Predict the initial focal plane} & \emph{Predict the initial focal plane}    \\ 
(c)  & Repeat z-stack creating for other tiles & \textbf{Repeat two-shot for other tiles}    \\ 
(d)  & Create a focus map& \textbf{Algorithm processing offline}   \\ 
(e)  & Shift platform for in-focus shooting & \textbf{Generate in-focus image directly} \\ 
\bottomrule
\label{table:workflow}
\vspace{-0.9cm}
\end{tabular}
\end{table}


\section{Conclusion}

In this paper, we presented a high-speed and high-throughput whole slide imaging system. Traditional autofocusing methods rely on repetitive mechanical adjustment to conduct refocusing, which is time-consuming. Instead, our scheme does not perform autofocusing during the process of tissue slide scanning, but recovers the in-focus image based on two-shot ones in an offline learning-based manner, as shown in Table \ref{table:workflow} . The proposed method is built upon the well-known U-Net, which is modified and extended such that it can work with two input images and yields a recovered in-focus image.
Experimental results demonstrate that our scheme achieves satisfactory performance on in-focus image recovery.


%



\section*{Acknowledgment}

The authors would like to thank Prof. G. Zheng and Dr. S. Jiang from UCOON for sharing the real measurements for WSI setup and beneficial discussions about \cite{jiang2018transform} as well as the following research.

\ifCLASSOPTIONcaptionsoff
  \newpage
\fi



%

{
\bibliographystyle{IEEEtran}
\bibliography{Autofocus}
}

\end{document}